%Paper: astro-ph/9502045
%From: aikman@dao.nrc.ca (Chris Aikman)
%Date: Tue, 7 Feb 95 11:28:04 PST

\magnification=\magstep1
\baselineskip 19pt
\centerline{IMAGING OF LOW REDSHIFT QSOs WITH WFPC2}
\vskip 20pt
\centerline{J.B.Hutchings\footnote{$^1$}{Observer with the NASA/ESA Hubble
Space Telescope through the Space Telescope Science Institute, which is
operated by AURA Inc., under contract NAS5-26555.}, S.C.Morris}
\centerline{Dominion Astrophysical Observatory, National Research Council
of Canada}
\centerline{5071 W.Saanich Rd, Victoria, B.C., V8X 4M6, Canada}
\vskip 20pt
\centerline{Abstract}
\vskip 5pt
Observations with the PC2 CCD of the Hubble Space Telescope are described
of two bright QSOs of redshift $\sim$0.3. 1403+434 is IR-bright and
radio-quiet,
and 2201+315 is radio-loud with extended structure.
Exposures were taken with the F702W and
F555W filters. The images are deconvolved on their own and combined with
0.5 arcsec ground-based images. Both host galaxies have the form and
luminosity of bright ellipticals, with nuclei of 1-2 times higher luminosity.
1403+434 is strongly interacting while 2201+315 may be in later stages of
a merger, both with a smaller companion. Both host galaxies have
compact knots and other small-scale peculiar features. Some general remarks
are made based on the total program sample of 6 QSOs.
\vfill\eject
\centerline{1. INTRODUCTION}
\vskip 10pt
This paper extends the work described by Hutchings et al (1994) on imaging
of selected low redshift QSOs with the corrected optics of the Hubble Space
Telescope. We report results on the last two QSOs in the program, and
discuss the total sample of six, of which 2 were observed before the HST
optical correction. The QSOs were selected from the extensive sample for
which we have CFHT imaging data, to cover a representative range of radio
luminosity, optical luminosity, and morphology as seen from the ground-based
data. Table 1 summarises the sample objects.

   The new observations were made with the F702W filter (4 x 140sec) and
F555W filter (500, 2 x 800, 1100 sec). The multiple reads allow easy
removal of significant cosmic ray events, and the range of exposure times
allows for correction for saturation effects at the QSO nuclei.
The major problems
in this work are the large dynamic range between the nucleus and the faint
host galaxy, and the shape of the PSF itself. It is clear in retrospect
that these difficulties need more careful attention in designing the
observations: ideally the pointing should be dithered at a sub-pixel level
to improve the sampling, and similar specific PSF observations should be
taken with a star at the same detector location.

    Nevertheless, the new data contain significant new information on
the morphology
and luminosity of bright knots and features in the host galaxies, and add
to our understanding of the activation of QSO nuclei. In what follows, we use
H$_0$=100, and q$_0$=0.5, although there is little senstivity to q$_0$ at
these redshifts.
\vskip 10pt
\centerline{2. 1403+434}
\vskip 10pt
   This z=0.32 QSO was discovered as an IRAS source by Low et al (1989).
These authors note that the spectrum shows a BAL Mg II line profile, and
that it is radio-quiet. Hutchings and Neff (1992) report CFHT imaging with
0.5 arcsec resolution, which suggests that the QSO is the nucleus of a round
and featureless galaxy which is in early interaction stages with a smaller
companion on a plunging orbit.

   The HST observations (Figure 1) support this conclusion. The tail consists
of two almost straight sections with a bend at a place of diffuse brighter flux
(knot B). The head of the tail (knot A) is extended almost tangentially
to the radius vector and there is no extra light closer in than this. The
outer end of the tail is somewhat brighter and fairly sharply bounded normal
to its length. It gets wider with increasing distance from the nucleus out
to this point. Two other faint knots (E,F) are seen within or superposed
on the opposite side of the host galaxy. They are seen as $\sim3\sigma$
detections
in both filters and are extended, although more compact
than a typical galaxy. Thus, they might be compact knots in the host galaxy.

    Restored images and PSF-subtracted images were made, using point spread
functions derived from star images and from the program Tiny Tim (Krist 1993).
The PSF in WFPC2 has structure which varies over the field because of variable
shadowing in the optical path and small optical differences across the
detectors. In addition, even the PC field is undersampled so that it is
almost impossible to match the pixel placement of PSF with the observation.
These conclusions were arrived at empirically, after many attempts at
restoration and subtraction.

   In the case of our QSOs, the PSF structure dominates the signal
over some 0.5 arcsec diameter, so that within this area we cannot
definitively isolate the host galaxy structure or luminosity profile.
Image restoration does improve the resolution of structure outside this
area, but in the outermost regions where the signal is weak, restoration tends
to enhance noise features. In no case (Figure 1) were we able to completely
remove the PSF structure, by restoration or by subtraction.

   Thus, the host galaxy luminosity profile was measured from the raw HST
image in which the main diffraction spikes are ignored or edited out,
as well as the interacting galaxy light. The
resulting profile does not fit an exponential, but does indicate a power
law (see Figure 2) in the radius range out to $\sim$3 arcsec. PSF
subtraction was performed from the profiles in Figure 2, and also in
flux-scaled images,
and on both cases the result is close to a straight line that fits the outer
part of the plots in Figure 2. This result is consistent with the
CFHT luminosity profile in Hutchings and Neff (1992), which extends over
larger radii, and for which a similar conclusion was drawn.
Thus, the host galaxy appears to be an elliptical with some minor
disturbance by its smaller merging companion.

   As in Hutchings et al (1994) and references therein, we have combined
the CFHT and HST images
with the Lucy-Hook algorithm, to enhance the resolution of the weak outer
signal. Figure 1 illustrates the results. The results are somewhat limited by
the large difference in resolution and signal between the datasets.
Nevertheless, the process improves the information on much of the structure,
especially
in the faint outer part of the tail. The failure of any restoration to
remove the PSF structure makes the restored images unreliable for luminosity
profiles. However, the combined restored images show similar profiles to the
raw data, but with significantly more scatter.

    Table 2 shows the photometry from this object and the features identified
in Figure 1. The nucleus is unusually
red for a QSO, consistent with its high IR luminosity. If the reddening is by
dust of a normal QSO nucleus, then there is $\sim$1.5 mag of extinction in the
observed R and $\sim$2 mag in observed V. This makes the nucleus quite luminous
and able to account for the observed IR luminosity by dust heating.
The light of the host galaxy estimated as described above, corresponds
to a luminous elliptical with M$_v$ = -22.3 $\pm$ 0.5, and fairly blue colour.
We discuss later the reliability of this estimate.

   The bright knot (A) at the head of the tail has a total luminosity of -18,
while the whole tail is about 1 magnitude brighter. Thus, if this is a
disrupted
galaxy, it has a luminosity similar to the LMC. The length of the tail is about
7.5 arcsec, which is about 20Kpc at the QSO redshift. The core of the bright
knot A at the head has M$_v\sim$-16, and is also apparently as red as the QSO
nucleus. All other colours are intermediate between this and the host galaxy
general colour: i.e. the knots and tail are redder than the main host galaxy
mass.

   The two faint knots (E,F)  on the opposite side have the brightness of
a 30 Dor type cluster, if they are associated with the QSO. They can just
be detected in the CFHT images.

   The most tantalising part of the image lies close to the nucleus. We have
removed the PSF by subtraction of a PSF image, and also by 180$^o$ rotation and
subtraction of the image from itself. This was done for both colours, and raw
and deconvolved images. In every case, there is a residual arc of luminosity
as shown in Figure 1(6), from about 2 o'clock to 5 o`clock
about 1 arcsec from the
nucleus. This arc might be a part of an elliptical ring that passes through
the head A of the tail, with the local extension of A forming part of this
ring.
The brightness cannot be measured very accurately but it appears to be
similar in colour to the main tail, and about 1 magnitude fainter.

   If this structure is real, then it implies either a disturbance deep within
the QSO host, or a double tail, with this second one superposed on the inner
part of the QSO host galaxy.
\vskip 10pt
\centerline{3. 2201+315}
\vskip 10pt
  This is a radio-loud QSO, also described in Hutchings and Neff (1991).
The redshift and optical luminosity are both similar to 1403+434. However,
the optical object has somewhat different morphology, and it is also a large
double-lobed radio source (see e.g. Hutchings and Gower 1984). The CFHT images
indicated an overall elliptical shape, extended normal to the radio axis,
but with a radial colour gradient (redder with increasing distance), and
some irregular knots within the host galaxy. It was hoped that the WFPC2
images would resolve some of the knots.

  The WFPC2 images had the same exposures as in 1403+434, so that the
F555W exposure is considerably deeper than the F702W. Both show weak
signal with little structure in the host galaxy. However, the 3 principal
features in the CFHT image are seen, and one more which is compact and
faint. These are labelled A to D in Figure 3. We also note that the two bright
nearby objects are unresolved and are presumably foreground stars. They have
redder colours ($\sim$1.2 mag) than any of the QSO features except for
the faint new knot A.

  Feature A is clearly extended in both HST and CFHT images. Features B and
C are very compact, but they are resolved. Feature D is so close to the
nucleus that its boundaries and structure are not well defined, but it
appears to be a large region which is brighter than the corresponding
area on the other side of the nucleus. Table 2 shows the measures made
on these features, based on both raw and restored images.

  Image restoration was done on the CFHT and WFPC2 data separately and
combined, and confirm the reality of the features we discuss. Restoration
tends to move signal into the nucleus in a way that may not be photometrically
correct. The ratio of nuclear to host galaxy flux in the raw images is
0.4 from CFHT and 0.8 from the WFPC. Restoration increases these to 1.0 and 3.2
respectively, and 1.2 in the combined restoration with equal weight to each.
A best estimate of 1.5 is adopted in Table 3. The total magnitudes are
close to those quoted in Hewitt and Burbidge (1993) and Hutchings and Neff
(1992), and the small differences are probably due to differing amounts of
the extended host galaxy that were measured.

   Figure 2 shows the luminosity profile from the edited raw HST image.
As with 1403+434, the restored image is not reliable for this measurement.
The inner 3 arcsec of 2201+315 clearly has a power law form and not an
exponential, as in 1403+434. The slopes are very similar, with some extra
light at $\sim$1 arcsec radius, as also seen in the CFHT data by Hutchings
and Neff (1992). The outer CFHT luminosity profile suggests some disturbance,
presumably from a merging event older than in 1403+434, which triggered the
radio source growth.

  The CFHT image shows that there are many faint small galaxies in the
field, which may be a more distant cluster. Their distribution, size, and
brightness are such that it is possible that feature A is one of them.
However, A is brighter, bigger, and bluer than most of them. Also, even
in the CFHT image, feature C appears more compact than these galaxies.
Thus, it seems reasonable to treat the features in Table 2 as associated
with the QSO.

   The region D is the brightest but it may reflect an off-centred nucleus
or central dust feature rather than a separate luminous entity. Its flux
is thus very uncertain and the region has no clear boundaries. The elongated
feature A is better defined and is the bluest part of the QSO environment,
and bluer than the nucleus. Region C is very compact and bright. Its size is
$\sim$4 pixels which corresponds to 500 pc.
It is conceivably the nucleus of a merging
galaxy, as in 1403+434, but there is no sign of a contiguous tail. The
feature B is also compact, but larger, and very red.

   Thus, we find that there are small regions and compact knots of
different colours, as suspected by Hutchings and Neff. Feature D may be
related to the overall shape of the galaxy, but otherwise none of the
features is related to any other, or to the overall galaxy, or to the
radio structure, which emerges perpendicular to the long axis of the
galaxy.
\vskip 10pt
\centerline{4. DISCUSSION}
\vskip 10pt
  The PC2 images are able to resolve compact features in the inner parts of
the QSO host galaxies. These are details of tidal tails, compact sites of
star-formation, or kpc-scale structure in the host galaxies. None of the
observed features is normal spiral structure, and in the sample of 6 in
Table 1, only 1229+204 shows
any kind of symmetrical structure - in that case a bar. In all 6 QSOs
we see compact blue knots which are consistent with being clusters of
several thousand young hot stars, perhaps like 30 Dor in the LMC.

  The objects were selected on the basis of known structure seen in the
pre-existing CFHT images, and so may be biased towards bright and peculiar
host galaxies. However, all objects appear
to be in the middle to late stages of merging with a smaller (M33 or LMC-size)
companion. The luminosity profiles are not simple and not characteristic
of an archetypical spiral or elliptical, which is
also consistent with the modelled
effects of merging. This is consistent with the discussion by Hutchings
and Neff (1992) based on 0.5 arcsec imaging of a larger and unbiased sample.

   The other WFPC2 radio-loud QSO in this program, 2141+175, has a featureless
host galaxy, with an excess of light on one side of the nucleus, similar
to region D seen here in 2201+315.
2141+175 has one faint compact knot in the host galaxy,
which is blue. 2141+175 has large tidal tails, not seen in 2201+315.
2141+175 is an unresolved radio source while 2201+315 is very large.
Thus, 2201+315 is likely to be an older source, in which there are no longer
major visible traces of a merger that triggered it. The two QSOs observed
with the aberrated HST (Table 1) are also radio-loud, and we cannot see the
same detail. However, they do contain knots of luminosity that are not
symmetrical or spiral in overall morphology.

   Of the four good HST host galaxy luminosity profiles,
only one shows an exponential shape. The one that runs counter to
`convention' is the (currently) radio-quiet IR-loud 1403+434. This is
apparently an elliptical in fairly early stages of merging, and with
significant
dust. Either of these circumstances may have suppressed formation of a central
radio source: Neff and Hutchings (1992) discuss this phenomenon in a sample
of IR-loud sources.

   We note that in a similar imaging program of bright low redshift QSOs,
Bahcall, Kirhakos, and Schneider (1994) report that all four of their
QSOs have low luminosity host galaxies. We have CFHT data on one of these
(0953+414) and also find in that object a faint host galaxy, in agreement with
their estimates. In previous
ground-based work (e.g. Hutchings, Janson and Neff 1989), we have found
a range of nuclear to host luminosity ratio of about 100.

  We have have two comments on this topic. First, in the cases where there
is overlap, the results from the CFHT programs are not
inconsistent with those from HST. This is encouraging, considering that
the ground-based data reach to significantly lower flux levels, in the
outer parts of the host galaxies, and also that the ground-based data have
considerably less host galaxy information close to the nucleus. Evidently,
the corrections made for these effects result in similar values for the
host galaxy luminosity.

   The second comment refers to empirical PSF-subtraction in the HST data.
In neither our nor the Bahcall et al subtractions is the PSF structure
eliminated, so that there is a significant uncertainty in the best
normalisation to use. (Note that data from the WFC detectors are less well
sampled by more than a factor 2.) The results shown by Bahcall et al
consistently show negative flux in the diffraction spikes. In
our subtractions, we regard this as overcorrected, so that they
may be underestimating the host galaxy signal as a result. Also, the faint
signal in the HST images can lead to underestimate of the host galaxy flux.

  Both our PSF-subtracted and our deconvolved images indicate host galaxies
which are luminous, as shown in Tables 2 and 3, and consistent with our
ground-based results. However, the complex PSF and weak signal make estimation
of the host galaxy flux and profile unreliable. The scaling of the PSF is
quite arbitrary and there is no definitive criterion to follow. We do not
regard the HST values for host galaxy flux as superior to ground-based.

  Thus, the results of Bahcall et al (1994) and those in this program appear
to represent different extremes in this regard, and possibly neither may
be used for general conclusions on host galaxy luminosities. It is not
generally true that QSO host galaxies are of low luminosity.

   It is clear that significant progress in QSO imaging with HST requires
attention to several things: observation of a PSF at the same position as
the QSO, with good signal; a large accumulated exposure (in short sections
to eliminate saturation and cosmic rays), sub-pixel dithering; and deep
ground-based images of the best possible resolution (0.5 arcsec or better).
\vfill\eject
\centerline{Captions to Figures}

1. 1403+434 images. 1,2 are CFHT raw and restored images in I band. 4 is
HST F555W raw image. 5 is combined CFHT + HST restored image, with
components marked (see Table 2). 3 and 6 are restored and PSF-subtracted
HST images on a larger scale. In panel 6 the shape of the nucleus A
can be seen, and the possible ring of extra flux on the opposite side
of the nucleus.

2. Azimuthally averaged luminosity profiles for the two QSOs from
F555W images, and the model PSF. The
innermost (saturated) pixels are not used, and the relative levels are
estimated from image subtraction. The actual scatter of points from the
PSF-subtracted image of 1403+434 is shown with an arbitrary displacement.
The PSF-subtracted profiles follow the lines defined by the QSO profiles
at radii larger than $\sim$0.5 arcsec. The two host galaxies have significant
flux and similar profiles.

3. 2201+315 images. 1,2 are CFHT raw and restored images. 4,5 are HST
raw images from F555W and F702W. Knots A, B, C  and the diffuse region
D are marked (see Table 3). 3 is the combined CFHT + HST restored image.
6 is the PSF-subtracted image, with the saturated central pixels masked out.
The extended flux around feature B can be seen.
\vfill\eject
\centerline{References.}

\noindent
Bahcall J.N., Kirkahos S., Schneider D.P. 1994, ApJ, 435, L11

\noindent

Hewitt A., Burbidge G., 1993, ApJS, 87, 451

\noindent
Hutchings J.B., Morris S.C., Gower A.C., Lister M.L. 1994, PASP, 106, 642

\noindent
Hutchings J.B., Holtzman J., Sparkes W.B., Morris S.C., Hanisch R.J., Mo J.,
1994, ApJ, 429, L1

\noindent
Hutchings J.B., Crabtree D., Neff S.G., Gower A.C. 1992, PASP, 104, 66

\noindent
Hutchings J.B., and Neff S.G. 1992, AJ, 104, 1

\noindent
Krist J., 1994, in ASP Conf Ser 52, Astronomical Data Analysis Software
and Systems II, ed R.J.Hanisch, R.J.V.Brissenden, and J.Barnes (San
Francisco: ASP).

\noindent
Low F.J., Cutri R.M., Kleinman S.G., and Huchra J.P. 1989, ApJ, 340, L1

\noindent
Neff S.G., and Hutchings J.B. 1992, AJ, 103, 1746.
\end

\magnification=\magstep1
\baselineskip 14pt
\nopagenumbers
\centerline{Table 1 QSOs observed with HST}
\vskip 10pt
\settabs 10\columns
\hrule
\vskip 3pt
\+QSO &~~~~Radio type &&m &z &Filter &Exp(sec) &&Comment\cr
\vskip 3pt
\hrule
\vskip 2pt
\hrule
\vskip 5pt
\+2305+187 &&RL &17.5 &0.31 &F702W &230 &Aberrated\cr
\+&&&&&F656N &1200 &Int galaxy with [O III]\cr
\vskip 5pt
\+1302-102 &&RL &14.9 &0.29 &F702W &800 &Aberrated\cr
\+&&&&&F555W &800 &Knots, assoc with radio?\cr
\vskip 5pt
\+1229+204 &&RQ &15.3 &0.06 &F702W &1400 &Merging barred spiral with\cr
\+&&&&&F606W &1400 &star-formation in tail\cr
\vskip 5pt
\+2141+175 &&RL &15.5 &0.21 &F702W &700 &Merging disk/elliptical\cr
\+&&&&&F606W &700 &Faint knot, jet\cr
\vskip 5pt
\+1404+434 &&RQ &16.5 &0.32 &F702W &700 &Merging red ell. galaxy\cr
\+&&&&&F555W &3200 &with bent tail\cr
\vskip 5pt
\+2201+315 &&RL &15.5 &0.30 &F702W &700 &Elliptical merger?\cr
\+&&&&&F555W &3200 &few knots\cr
\vskip 3pt
\hrule
\vskip 8pt
All QSOs also have 0.5 arcsec imaging with CFHT in V and I band.
\end

\nopagenumbers
\baselineskip 16pt
\magnification=\magstep1
\centerline{Table 2. 1403+434}
\vskip 10pt
\settabs 8\columns
\hrule
\vskip 3pt
\+Feature &&F702W &F555W &`V-R' &M$_v$\cr
\+&&`R' mag &`V' mag\cr
\vskip 3pt
\hrule
\vskip 2pt
\hrule
\vskip 10pt
\+Whole &&15.6 &16.6 &1.0 &-23.7\cr
\+Nucleus &&16.1 &16.9 &0.8 &-23.4 (-25.3$^a$)\cr
\+Host &&16.7 &17.9 &1.2 &-22.3\cr
\+$\Sigma$A &&21.1 &22.0 &0.9 &-18\cr
\+A nuc &&22.6 &23.8 &(1.2) &-16\cr
\+B &&22.3 &22.9 &0.6 &-17\cr
\+C &&23.3 &23.8 &0.5 &-16\cr
\+B+C+D &&20.6 &21.2 &0.6 &-19\cr
\+Inner arc &&(21.7) &(22.5) &(0.8) &(-17)\cr
\+E &&(25.2) &25.5 &(0.3) &(-14)\cr
\+F &&(25.2) &25.9 &(0.7) &(-14)\cr
\vskip 3pt
\+Nuc/Host &&1.7 &2.5\cr
\vskip 3pt
\hrule
\vskip 10pt
$^a$ Value dereddened for E$_{B-V}$=0.7

Values in parentheses are less certain

Features A - F identified in Figure 1
\end

\magnification=\magstep1
\nopagenumbers
\baselineskip 16pt
\centerline{Table 3. 2201+315}
\vskip 10pt
\settabs 8\columns
\hrule
\vskip 3pt
\+Feature &&F702W &F555W &`V-R' &M$_v$\cr
\+&&`R' mag &`V' mag\cr
\vskip 3pt
\hrule
\vskip 2pt
\hrule
\vskip 10pt
\+Whole &&15.3 &15.8 &0.5 &-24.1\cr
\+Nucleus &&15.9 &16.2 &0.3 &-23.7\cr
\+Host &&(16.3) &(17.0) &(0.7) &(-23)\cr
\+A &&24.4 &27.2 &2.8 &-12.7\cr
\+B &&23.5 &23.6 &0.1 &-16.3\cr
\+C &&23.5 &23.9 &0.4 &-16.0\cr
\+D &&- &(22.2) &- &(-17.7)\cr
\vskip 3pt
\+Nuc/Host &&(1.5) &(2.0)\cr
\vskip 3pt
\hrule
\vskip 10pt
Values in parentheses are less certain

Features A - D identified in Figure 3
\end